\documentclass[
  reprint,
  superscriptaddress,
  amsmath,amssymb,
  aps,
  longbibliography,
  floatfix
]{revtex4-2}

\usepackage[T1]{fontenc}
\usepackage[utf8]{inputenc}
\usepackage{graphicx}
\usepackage{bm}
\usepackage{xurl}
\usepackage[hidelinks]{hyperref}
\hypersetup{
  pdftitle={Collective Hall quantization in 3D Chern vector systems},
  pdfauthor={Zhi-Qiang Zhang, Shu-Guang Cheng, Hongfang Liu, Hailong Li, Hua Jiang, X. C. Xie},
  pdfsubject={Research article in APS REVTeX 4.2 format}
}

\begin{document}

\title{Collective Hall quantization in 3D Chern vector systems}

\author{Zhi-Qiang Zhang}
\affiliation{Interdisciplinary Center for Theoretical Physics and Information Sciences, Fudan University, Shanghai 200433, China}
\affiliation{School of Physical Science and Technology, Soochow University, Suzhou 215006, China}

\author{Shu-Guang Cheng}
\affiliation{Department of Physics, Northwest University, Xi'an 710069, China}

\author{Hongfang Liu}
\affiliation{School of Physical Science and Technology, Soochow University, Suzhou 215006, China}

\author{Hailong Li}
\email{hailongli@fudan.edu.cn}
\affiliation{Interdisciplinary Center for Theoretical Physics and Information Sciences, Fudan University, Shanghai 200433, China}

\author{Hua Jiang}
\email{jianghuaphy@fudan.edu.cn}
\affiliation{Interdisciplinary Center for Theoretical Physics and Information Sciences, Fudan University, Shanghai 200433, China}

\author{X. C. Xie}
\affiliation{Interdisciplinary Center for Theoretical Physics and Information Sciences, Fudan University, Shanghai 200433, China}
\affiliation{International Center for Quantum Materials, School of Physics, Peking University, Beijing 100871, China}
\affiliation{Hefei National Laboratory, Hefei 230088, China}

\begin{abstract}
Understanding the three-dimensional (3D) quantum Hall effect,
characterized by a Chern vector rather than a single Chern number,
has been a central pursuit in topological physics.
While quantized Hall transport is established for the special case
with one nonzero Chern component, recent realizations of more
general Chern-vector states raise a fundamental question:
how does such topology protect quantized transport?  Here, we predict a distinct 3D quantized Hall effect protected
by $\mathbf{C}=(0,m,n)$.
For samples with dimensions $L_y$ and $L_z$, global topology
enforces collective quantization of $mL_y+nL_z$, despite the
fragility of individual edge modes to disorder.
This reveals a novel transport regime in which robust quantization
emerges as a holistic property of a unified chiral network,
rather than of its constituent modes.
Crucially, this collective response manifests as a quantized
Hall conductance $G_{xy}=\pm(mL_y+nL_z)e^2/h$,
with a distinctive dependence on both Chern-vector components
and sample dimensions.
We propose candidate materials to facilitate experimental observation.
\end{abstract}

\keywords{Chern vector, 3D quantum Hall effect, collective quantization, chiral edge network, topological transport}
\maketitle

\section{Introduction}

\begin{figure*}[!t]

    \centering
    \includegraphics[width=\textwidth]{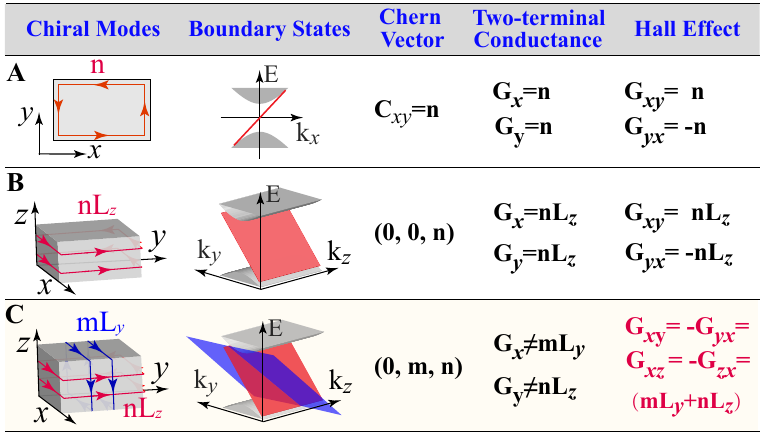}

    \caption{Chiral modes, front surface state spectra, conductances and Hall effect from Chern-number insulators to Chern-vector insulators. (A) 2D samples with $\mathcal{C}_{xy}\!=\!n$;
(B) 3D samples with a single nonzero Chern component $\textbf{C}\!=\!(0,0,n)$;
(C) 3D samples with multiple nonzero Chern components $\textbf{C}\!=\!(0,m,n)$, with $t_d=0$ for the schematic spectra. The sample sizes are $L_\alpha$ with $\alpha\!\in\! x,y,z$. $G_\alpha$ is the two-terminal conductance along $\alpha$ direction. To simplify the analysis, we set $m \!=\! n \!=\! 1$ and $e^2/h \!=\! 1$ by default.
	\label{f1}}
\end{figure*}

The two-dimensional (2D) quantized Hall effect (QHE) \cite{zongshu1,zongshu2,shen1,QH1,QH2,QH3,QAH1,QAH2,QAH3,QAH6,QAH7,QAH8,QAH9,QAH10,QAH11,QAH12,Mn5,3DQHCDW,3DQHE1,3DQHE2} has long been a cornerstone of  topological physics, establishing a  profound connection between the Chern number $\mathcal{C}_{xy}$,  chiral edge modes, and the  precisely quantized Hall conductance $G_{xy}\!=\!\mathcal{C}_{xy}e^2/h$ \cite{QH2,QH3,QAH1}, as shown in Fig.~\ref{f1}(A).
Recently, interest has expanded to three-dimensional (3D) systems where various Hall responses have been proposed and experimentally verified \cite{QAH11,QAH12,Mn5,3DQHCDW,3DQHE1,3DQHE2,3DQHE3,3DQHE4,3DQHE5,3DHE1,3DHE2,3DHE3,3DHE4,3DHE5,Wu2026FermiArcs,PhysRevLett.133.126602}.
Generalizing the scalar Chern number to 3D systems yields the Chern vector $\mathbf{C}\!=\!(\mathcal{C}_{yz},\mathcal{C}_{xz},\mathcal{C}_{xy})$ \cite{Chernvector,Chernvector2,Chernvector3,Chernvector4}, which governs both the spatial configuration and transport of chiral boundary states.
For example, when $\mathbf{C}=(0,0,n)$, as shown in Fig.~\ref{f1}(B), each of the $L_z$ structural periods along the $z$-direction contributes a Chern number $n$, resulting in a quantized Hall conductance $G_{xy} = nL_z e^2/h$.
While this framework has successfully underpinned research on tunable quantum anomalous Hall effects and led to significant experimental milestones \cite{QAH11,QAH12,Mn5,3DQHCDW}, Chern vectors with multiple nonzero components---which host far more intricate topological boundary states---remain largely unexplored in quantum transport.

Very recently, 3D topological states characterized by Chern vectors $\mathbf{C}\!=\!(0,m,n)$ and their exotic surface states have been experimentally realized \cite{Chernvector,Chernvector2,Chernvector4}. However, existing studies have primarily focused on their static band dispersions. How quantum transport is fundamentally governed by a multi-component Chern vector remains an unresolved and pressing challenge.
Crucially, a Chern vector of the form $\mathbf{C}\!=\!(0, m, n)$ manifests as an interconnected chiral network [Fig.~\ref{f1}(C)], where complex loops and cross-surface links induce substantial scattering between opposite surfaces. This behavior fundamentally departs from that of conventional 2D Chern insulators \cite{QAH2} or simple 3D stacks with $\mathbf{C}\!=\!(0,0,\mathcal{C}_{xy})$ \cite{QAH11}. Under such inter-surface scattering, the individual chiral modes---though nominally protected by Chern numbers---become ``fragile'': they no longer sustain the backscattering-free transport that characterizes the 2D paradigm \cite{buttiker}. Consequently, standard frameworks, such as the layer-based TKNN formula \cite{3DTKNN1,3DTKNN2}, fail to capture the transport physics, raising a fundamental question: Can any quantized transport signature survive when local edge-state protection is compromised by global network complexity?

Here, we demonstrate that a 3D quantized Hall effect is uniquely protected by the Chern vector $\mathbf{C}\!=\!(0,m,n)$ through an emergent mechanism of collective quantization.
We reveal that although the transmission of individual edge states deviates from quantization due to inter-surface scattering, the total transmission coefficient of the unified chiral network remains robustly quantized at $(mL_y+nL_z)$.  This collective immunity transcends the local fragility of constituent modes, revealing a  shift from \textit{individual} to \textit{holistic} topological protection.
Applying the Landauer-B{\"u}ttiker formalism \cite{LB1}, we show that this  topological protection underpins distinct multi-terminal responses measurable in standard Hall-bar setups, leading to four quantized Hall conductances: $G_{xy}\!=\!G_{xz}\!=\!-G_{yx}\!=\!-G_{zx}\!=\!(mL_y\!+\!nL_z)e^2/h$.
These conductances exhibit a definitive scaling with the sample dimensions $(L_y, L_z)$, directly reflecting the intrinsic 3D nature of the effect. Finally, we propose realistic solid-state candidate materials to facilitate experimental realization. Our work bridges the gap between formal 3D topological invariants and measurable quantized transport.

\begin{figure*}[!t]

    \centering       
    \includegraphics[width=\textwidth]{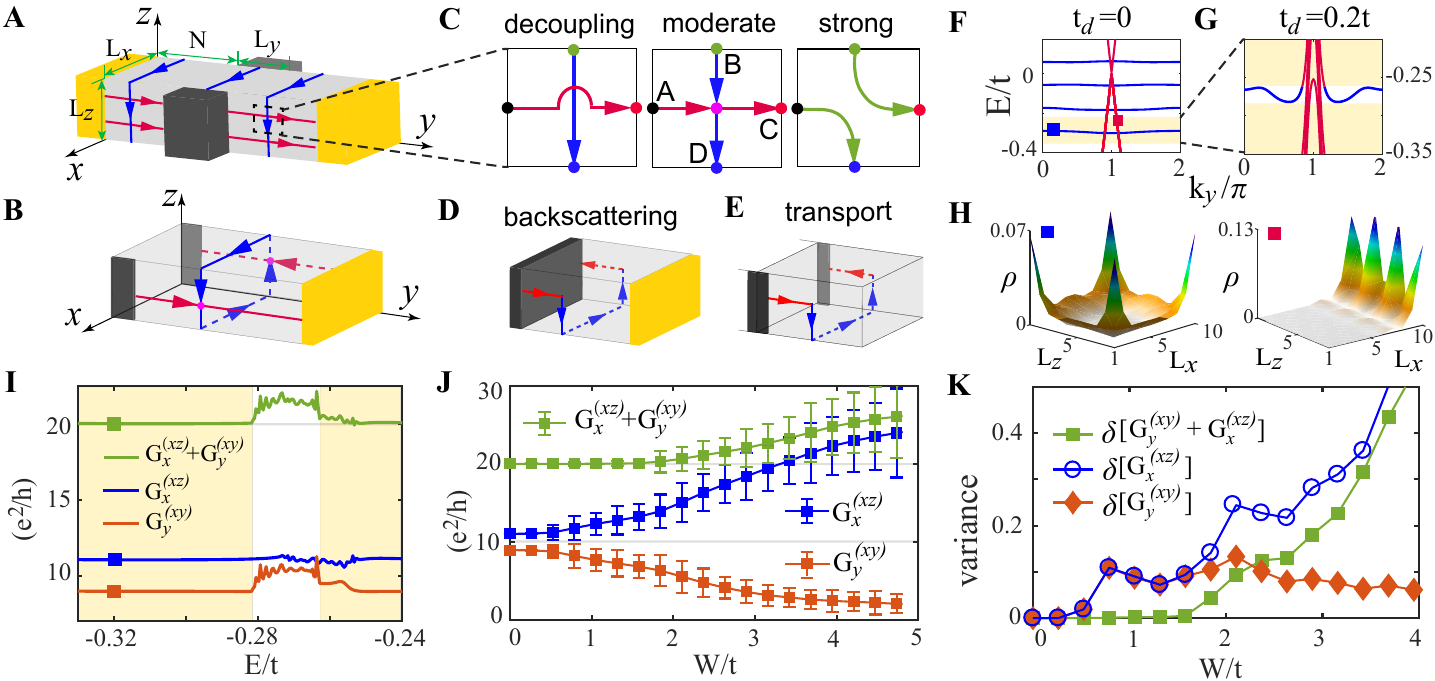}

    \caption{Chiral surface states and quantized transport in a Chern-vector insulator.
(A) Schematic illustration of the chiral surface states in a system with Chern vector $\mathbf{C}=(0,1,1)$. The sample details are indicated in the figure.
(B) Half of the sample in (A), where solid (dashed) lines denote the visible (hidden) chiral modes.
(C) Enlarged views of a typical chiral intersection in (A) for different inter-mode coupling strengths $t_d$.
(D),(E) Chiral network transport for various lead configurations in (B).
(F),(G) Energy spectra $E$ versus $k_y$ with and without inter-mode coupling $t_d$, respectively.
(H) Typical spatial distributions of the eigenstates, $\rho=|\psi_n|^2$, for the blue and red modes indicated in (F).
(I) Two-terminal conductance $G_\gamma^{(\alpha\beta)}$ as a function of Fermi energy $E$.
We define $G_\gamma^{(\alpha\beta)}$ as the conductance of chiral modes localized at the $\alpha$-$\beta$ boundaries, where the current flows along the $\gamma$ axis, for $\alpha, \beta, \gamma \in \{x, y, z\}$.
(J) $G_\gamma^{(\alpha\beta)}$ as a function of disorder strength $W$ at Fermi energy $E=-0.32t$.
(K)  shows the variance of the data in (J).
The parameters are $L_x=L_y=L_z=10$ and $N=15$.}
	\label{f2}
\end{figure*}

\section{Model}
Our starting point is a minimal tight-binding model composed of two
coupled 3D Chern-insulator sectors [see Appendix  and Supplementary
Material (SM) Note~0 \cite{SM} for details],

\begin{equation}\label{Eq1}
\begin{aligned}
H={}&\begin{pmatrix}\mathcal{H}_{xyz}&0\\0&\mathcal{H}_{xzy}\end{pmatrix}\\
&+\sum_n\begin{pmatrix}0&T_d\\T_d&0\end{pmatrix}c_n^\dagger c_n.
\end{aligned}
\end{equation}

Here, $\mathcal{H}_{xyz}$ describes a 3D Chern insulator characterized by
the Chern vector
$\mathbf{C}_{xyz}
=
(\mathcal{C}_{yz},\mathcal{C}_{xz},\mathcal{C}_{xy})
=
(0,0,1)$,
\cite{QAH1,QAH2,QAH3,QAH6,QAH7,QAH8,QAH9,QAH10,QAH11,QAH12}
whereas $\mathcal{H}_{xzy}$ is obtained by exchanging the roles of the
$y$ and $z$ directions and is characterized by
$\mathbf{C}_{xzy} = (0,1,0)$. The explicit real-space and momentum-space forms of
$\mathcal{H}_{xyz}$ and $\mathcal{H}_{xzy}$ are given in the Appendix.

According to the bulk-boundary correspondence,
$\mathcal{H}_{xyz}$ gives rise to red chiral modes associated with the
$x$-$y$ boundaries, while $\mathcal{H}_{xzy}$ produces blue chiral modes
associated with the $x$-$z$ boundaries, as illustrated in
Fig.~\ref{f2}(A). In the decoupled limit, their combination therefore
realizes the composite Chern vector $\mathbf{C}=
(0,1,1)$.

The two sectors are coupled through the hopping matrix
$T_d=t_d(\mathbf{I}_{2\times2}+\tau_x)$,
where $t_d$ denotes the inter-sector coupling strength. A finite $t_d$
hybridizes the intersecting chiral surface modes and modifies their
connectivity without closing the bulk gap in the parameter regime
considered here [see Appendix and SM Note~1].

To examine the robustness of the transport properties, Anderson
disorder is included through
$H_w=\sum_n\varepsilon_n c_n^\dagger c_n$, where $\varepsilon_n$ is
uniformly distributed within $[-W/2,W/2]$, with $W$ denoting the
disorder strength \cite{Anderson0,Anderson1,Anderson2}.

\section{Chiral network configuration and terminal responses}

We first investigate the influence of coupling $t_d$ on the chiral mode configurations. As shown in Fig.~\ref{f2}(A) and (C), the surface networks exhibit coupling-dependent scattering between red and blue chiral channels with probability $P(t_d)$, spanning three limits: (i) the decoupling limit where $P_{t_d=0}=0$; (ii) the moderate coupling regime with finite $P_{t_d\neq0}\neq0$; and (iii) the strong coupling limit where $P_{t_d\rightarrow\infty}=1$, which reconfigures the network geometry.

Crucially, the bulk Chern vector $\mathbf{C}=(0,1,1)$ remains invariant under varying $t_d$ [see SM. Note~1].
Despite these structural variations [a feature absent in the $\mathbf{C}=(0,0,1)$ case], all configurations here obey a universal unidirectional scattering rule: tunneling occurs strictly from $\mathrm{A}/\mathrm{B}$ to $\mathrm{C}/\mathrm{D}$ [Fig.~\ref{f2}(C)]. As this rule is consistently captured by configuration (ii), this regime provides a unified framework to describe transport across all coupling strengths, as governed by $\mathbf{C}=(0,1,1)$.

To characterize the topology, we examine the two-terminal conductance $G^{(\alpha\beta)}_\gamma$. Here, we introduce the notation $G^{(\alpha\beta)}_\gamma$ to denote the two terminal conductance in the $\gamma$-direction, originating from the chiral modes located on the $\alpha$-$\beta$ boundaries. For example, in the conventional case [Fig.~\ref{f1}(B)], the conductance is topologically protected and determined solely by the planar Chern number with $G^{(xy)}_y = \mathcal{C}_{xy}L_z$.

However,  for a general Chern vector $\mathbf{C}=(0, \mathcal{C}_{xz}, \mathcal{C}_{xy})$, the two-terminal responses lose this topological protection. Although the components of the Chern vector remain invariant as per the TKNN formula, the non-zero $t_d$ induces scattering within the chiral network, fundamentally altering the transport.
In the $y$ direction, the red modes exhibit traveling features that dominate the conductance, whereas the blue modes are nearly localized in flat bands [Fig. \ref{f2}(F)].
However, scattering between these modes partially gaps out the spectrum [Fig. \ref{f2}(G)], causing $G^{(xy)}_y$ to deviate from the quantized value $\mathcal{C}_{xy}L_z$. For instance, in both clean [Fig. \ref{f2}(I)] and disordered [Fig. \ref{f2}(J)] cases, $G^{(xy)}_y$ falls below the expected value of $\mathcal{C}_{xy}L_z=10$. Intuitively, this reduction, expressed as $G^{(xy)}_y=[\mathcal{C}_{xy}-P_1(t_d)]L_z$, originates from backscattering induced by the coupling between chiral modes on opposite surfaces [Fig. \ref{f2}(D)].

In the $x$-direction, $G^{(xz)}_x$ similarly deviates from the quantized value $\mathcal{C}_{xz}L_y$, but notably shifts toward a higher value in both clean and disordered cases [Figs. \ref{f2}(I)--(J)]. This enhancement, described by $G^{(xz)}_x = \mathcal{C}_{xz}L_y + P_2(t_d)L_z$, stems from additional transport channels facilitated by inter-mode scattering [Fig. \ref{f2}(E)]. Because these typical two-terminal responses $G_y^{(xy)}$ and $G_x^{(xz)}$ lack topological protection, they cannot directly reflect the intrinsic $(0,1,1)$ topology. This discrepancy necessitates an alternative mechanism to explain the robust, quantized transport observed in the system.

Remarkably, the scattering processes in Fig. \ref{f2}(B) that decrease $G^{(xy)}_y$ [Fig. \ref{f2}(D)] and increase $G^{(xz)}_x$ [Fig. \ref{f2}(E)] are physically identical, inherently ensuring charge conservation [See SM. Note 3 for details]. This suggests that the summation $G^{(xz)}_x\!+\!G^{(xy)}_y$ remains topologically protected provided $P_1(t_d)=P_2(t_d)$, as the scattering-induced deviations exactly cancel each other out:
\begin{equation}\label{conservation}
\begin{aligned}
G^{(xz)}_x+G^{(xy)}_y
&=[\mathcal{C}_{xz}L_y+P_2(t_d)L_z]\\
&\quad+[\mathcal{C}_{xy}L_z-P_1(t_d)L_z]\\
&=\mathcal{C}_{xz}L_y+\mathcal{C}_{xy}L_z.
\end{aligned}
\end{equation}
This quantized sum defines a total Chern vector $\mathbf{C}_{\text{t}} = (0, \mathcal{C}_{xz}L_y, \mathcal{C}_{xy}L_z)$, serving as a robust measure of the aggregate transport channels within a typical sample.

To verify this conclusion, Fig. \ref{f2}(I) shows numerical results with $\mathbf{C}_{\text{t}} = (0, 10, 10)$ demonstrating that $G^{(xz)}_x\!+\!G^{(xy)}_y$ is quantized at $\mathcal{C}_{xz}L_y+\mathcal{C}_{xy}L_z=20$ within the shaded energy interval \cite{footnote1}.
 Furthermore, the robustness of this topological quantity is confirmed in Figs. \ref{f2}(J) and \ref{f2}(K). $G^{(xz)}_x\!+\!G^{(xy)}_y$ remains quantized with vanishing fluctuations for significant disorder strengths and varying $t_d$ [refer to SM. Note 4]. This stands in sharp contrast to the fragile nature of the individual components $G^{(xz)}_x$ and $G^{(xy)}_y$, which fluctuate considerably.

 Correspondingly, the total $x$-direction conductance $G_x = G^{(xz)}_x + G^{(xy)}_x$ emerges as a similarly robust topological response.

Significantly, we find that the quantized sum $(\mathcal{C}_{xz}L_y + \mathcal{C}_{xy}L_z)$ is equivalent to the global Hall conductance, which is directly measurable in a standard Hall-bar geometry. To demonstrate this, we model the quantum transport of a sample with $\mathbf{C} = (0, 1, 1)$ using the Landauer-B{\"u}ttiker formalism. For a multi-terminal device, the relationship between current and voltage is governed by \cite{LB1}:
\begin{equation}\label{LBequation}
I_i=\frac{e^2}{h}\sum_{j\neq i}T_{ij}(V_i-V_j).
\end{equation}
$V_i$ represents the voltage at the $i$-th lead, $I_i$ denotes the current flowing out of the $i$-th lead into the sample, and $T_{ij}$ is the transmission coefficient from the $j$-th to the $i$-th lead.

\begin{figure*}[!t]

    \centering       
    \includegraphics[width=0.82\textwidth]{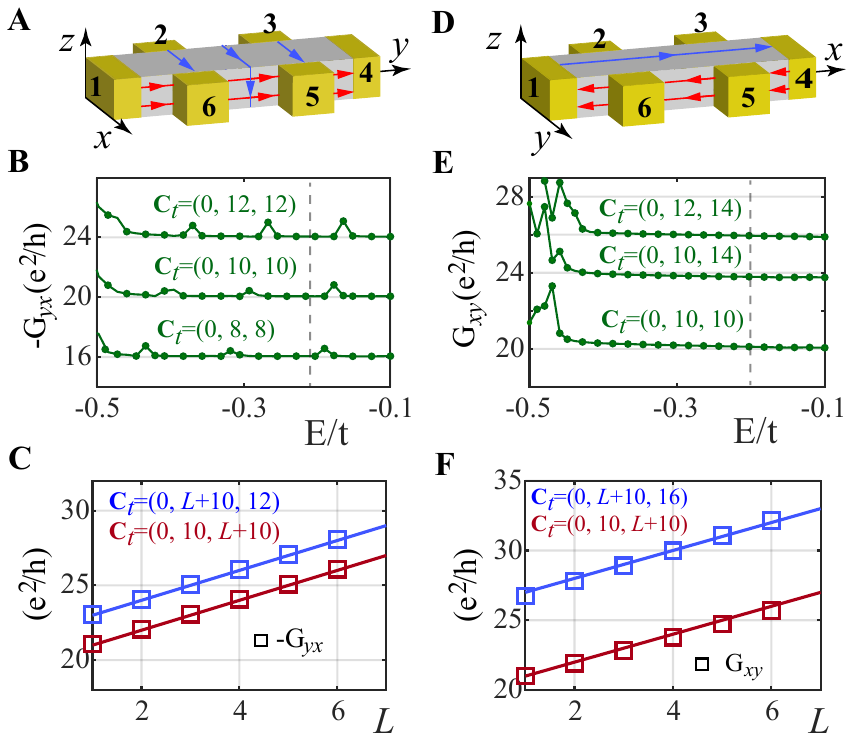}

    \caption{Hall bar device measurement of quantized Hall conductance.
    (A) Schematic plot of Hall bar with current along the $y$-direction.
(B) and (C) the Hall conductances versus Fermi energy $E$ and $\textbf{C}_t$ for the setup in (A). The total Chern vector $\textbf{C}_t=(0, \mathcal{C}_{xz}L_y, \mathcal{C}_{xy}L_z)$ is marked in the figure with $t_d=0.1t$. The solid lines plot $(\mathcal{C}_{xz}L_y\!+\!\mathcal{C}_{xy}L_z)$ versus $L$. (D)-(F) are the same as those in (A)-(C), excepting currents along the $x$-direction. We set $E\!=\!-0.2t$ for (C) and (F).}
	\label{f3}
\end{figure*}

We first investigate transport in a Hall bar oriented along the $y$-direction, characterized by the total Chern vector $\mathbf{C}_{\text{t}} = (0, \mathcal{C}_{xz}L_y, \mathcal{C}_{xy}L_z)$. For the standard six-terminal geometry shown in Fig. \ref{f3}(A), current is injected between leads 1 and 4, while leads 2, 3, 5, and 6 serve as voltage probes. The interconnected chiral modes yield the following non-zero transmission coefficients:
\begin{equation}
\begin{split}
T_{j,j+1}&=G^{(xy)}_y,~T_{61}=G^{(xy)}_y,\\
T_{26/35}&=G^{(xz)}_x,~T_{62/53}=G^{(xz)}_x
\end{split}
\end{equation}
where $j\in[1,2,3,4,5]$. By solving the Eq.~(\ref{LBequation}) yields the quantized Hall conductance between leads 2 and 6 [refer to SM. Note 5 for details]
\begin{equation}\label{Hallyx}
\begin{aligned}
G_{yx}&=I_y/(V_2-V_6)\\
&=-(\mathcal{C}_{xz}L_y+\mathcal{C}_{xy}L_z),
\end{aligned}
\end{equation}
where $I_y$ denotes the current along the $y$-direction.

Due to the 3D nature of the Chern vector, the Hall-bar response can also be measured along the $x$-direction [Fig. \ref{f3}(D)]. In this orientation, the Hall conductance remains quantized as [see SM. Note 5 for details]:
\begin{equation}\label{Hallxy}
\begin{aligned}
G_{xy}&=I_x/(V_2-V_6)\\
&=\mathcal{C}_{xz}L_y+\mathcal{C}_{xy}L_z.
\end{aligned}
\end{equation}
The validity of Eq.~(\ref{Hallyx}) and Eq.~(\ref{Hallxy}) is verified by numerical calculations using the non-equilibrium Green's function technique \cite{LB1}. As shown in Fig. \ref{f3}(B), the Hall conductance $G_{yx}$ aligns with the quantized value $(\mathcal{C}_{xz}L_y+\mathcal{C}_{xy}L_z)$ across varying sample sizes. A similar agreement is observed for $G_{xy}$ in Fig. \ref{f3}(E). The robustness of these Hall responses against disorder is further confirmed in SM. Note 6.

Unlike the single Chern number dependence of $G_{xy}\!=\!nL_z$, typically observed for $\textbf{C}_t\!=\!(0,0,nL_z)$ [Fig. \ref{f1}(B)], the 3D quantized Hall conductance of $\textbf{C}_t\!=\!(0,mL_y,nL_z)$ exhibits double Chern number components dependence and can be observed on different directions. Specifically, both $G_{yx}$ and $G_{xy}$ exhibit a joint dependence on Chern numbers $mL_y$ and $nL_z$ [Figs. \ref{f3}(C) and (F)], serving as a distinctive experimental fingerprint of the collective 3D quantized Hall effect.

\section{Intrinsic network topology vs. principal axis rotation}
To elucidate the topological essence of the general Chern vector $\mathbf{C} = (0, m, n)$, it is instructive to contrast our framework with the conventional $(0, 0, q)$ case obtained via coordinate rotation. While a Chern vector is mathematically reducible to a principal axis, such a mapping obscures the rich surface physics emerging from the internal coupling within the chiral network.

\begin{figure*}[!t]
    \centering

\includegraphics[width=\textwidth]{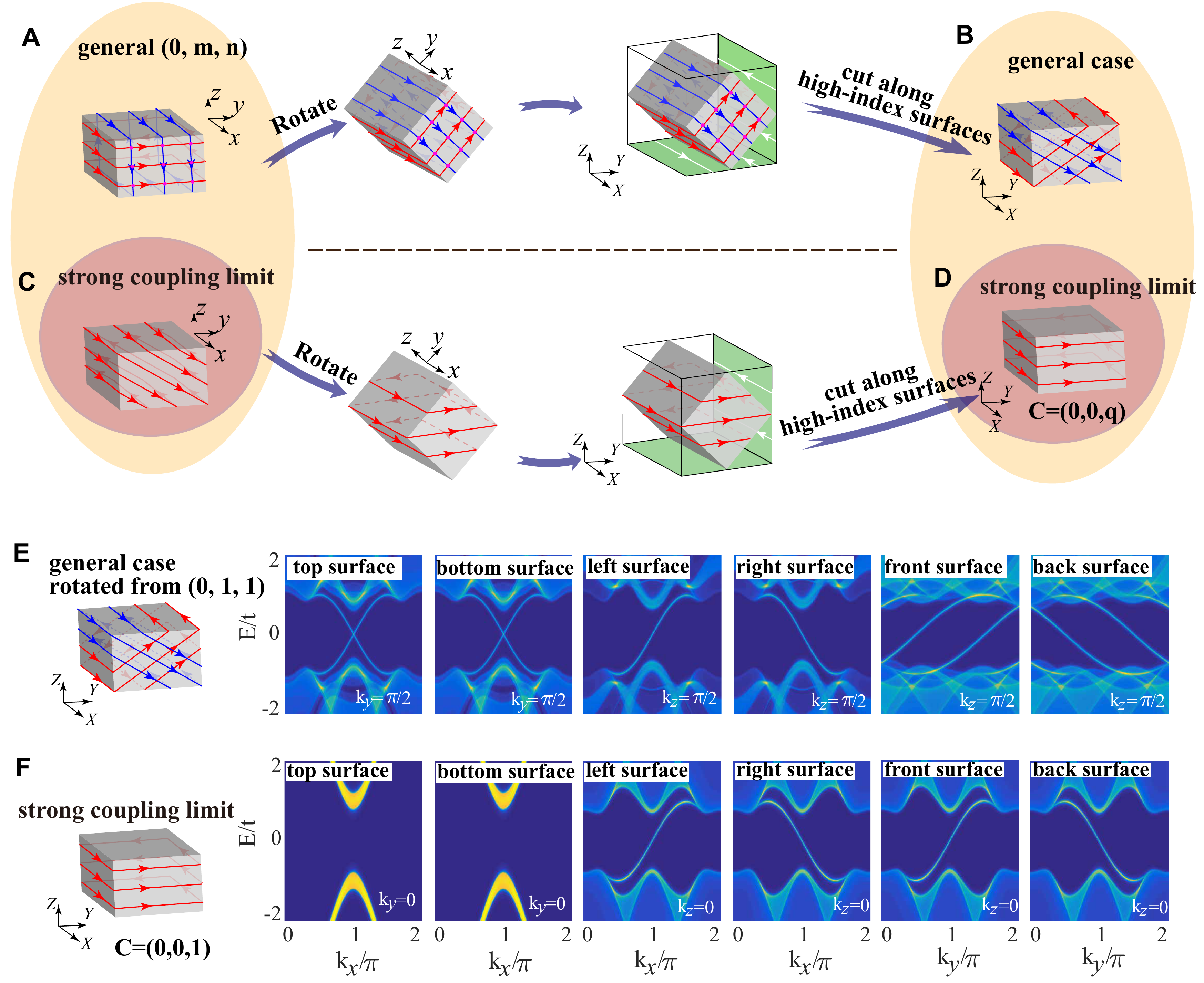}

	\caption{Intrinsic 3D chiral network topology beyond stacked 2D QAH physics.
	Schematic of surface states for a 3D Chern insulator with a general Chern vector $\mathbf{C} = (0, m, n)$ in the general case (A) and the strong-coupling limit (C).
	(B) and (D) show surface configurations after coordinate rotation and cutting along high-index surfaces. In the general regime (B), the chiral network persists, maintaining a complex connectivity that cannot be reduced to simple 2D physics. In contrast, the strong-coupling limit (D) yields a configuration equivalent to a stacking of 2D QAH insulators with $\mathbf{C} = (0, 0, q)$. (E) and (F) give their typical surface spectra.}
	\label{Fig5}
\end{figure*}

As illustrated in Fig. \ref{Fig5}, the global connectivity of the chiral network persists even when the system is cut along high-index surfaces (Fig. \ref{Fig5}(A)), maintaining a complex topology fundamentally distinct from the decoupled stacking limit (Fig. \ref{Fig5}(C)). This inequivalence is further evidenced by the surface spectra: in the general regime, unique surface states emerge on both top and bottom surfaces (Fig. \ref{Fig5}(E)), a spectral hallmark of the 3D chiral network. In the strong-coupling limit, however, these features vanish as the system collapses into a simple stacking of 2D QAH layers (Fig. \ref{Fig5}(F)). Generally, the transport phenomena and quantized Hall responses observed in this work originate from the intrinsic network topology, providing a theoretical breadth that encompasses but significantly extends beyond the traditional rotation-and-stacking schemes.

In general, the standard stacked-layer picture is recovered only in the singular strong-coupling limit, where the system's intrinsic complexity effectively decouples.
However, this limit represents a constrained regime that is not representative of the general physics.
In the broader regime of intermediate coupling, the system manifests a 3D collective state whose topology is governed by global network connectivity.
This global network architecture accounts for the full variety of chiral mode configurations, that remains fundamentally inaccessible within the spatial rotation of the $(0,0,q)$ limit.
Consequently, our generalized framework establishes a comprehensive theoretical platform, while the rotated $(0,0,q)$ case represents a limiting simplification rather than a full realization of the general physics.

\begin{figure*}[!t]

    \centering       
    \includegraphics[width=0.82\textwidth]{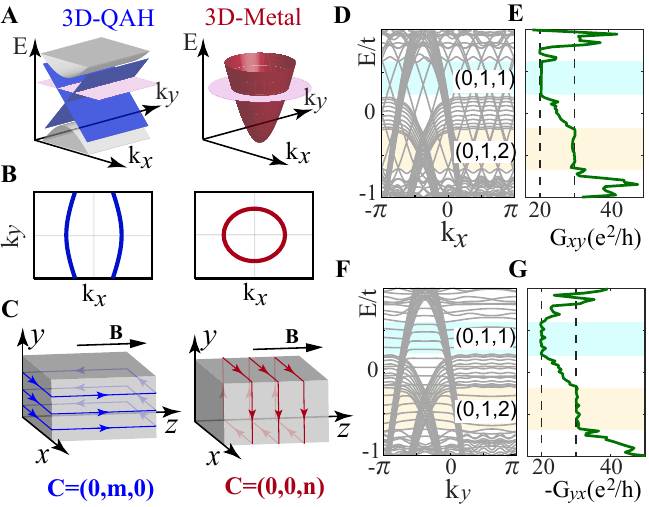}

    \caption{Quantized Hall response in Chern-vector-protected 3D quantum Hall insulators.
(A) Schematic band structures for a 3D QAH and a 3D metallic phase.
The pink planes indicate the constant-energy surfaces.
(B) Corresponding Fermi surfaces for the spectra shown in (A).
(C) Surface states associated with (A) under an applied magnetic field.
(D),(F) Numerically calculated band spectra at magnetic flux $\phi=0.25\pi$.
(E) and (G) are Hall conductances of (D) and (F), respectively.
The sample sizes are $L_x\!=\!L_y\!=\!L_z\!=\!10$. The other parameters are fixed at $ t_x = t_y = t $, $ \varepsilon_0 = -2t $, $ t_z = 0.1t $, and $t_d\!=\!0.1t$}
   \label{f4}
\end{figure*}

\section{Candidate systems and materials}
The universality of this network-based framework allows for its realization across diverse topological platforms. Because its transport properties stem from intrinsic network topology rather than specific structural orientation, the framework transcends the constraints of geometric rotation models. To demonstrate this versatility, we consider a representative hybrid system where 3D Chern insulator bands coexists with metallic bands-a configuration prevalent in many topological materials. This system, governed by the non-local connectivity of the network, is described by the Hamiltonian:
\begin{equation}\label{Eq2}
\begin{aligned}
H={}&\begin{pmatrix}
\mathcal{H}_{\text{metal}}(\mathbf{B}) & 0\\
0 & \mathcal{H}_{xzy}(\mathbf{B})
\end{pmatrix}\\
&+\sum_n\begin{pmatrix}0&t_d\\t_d&0\end{pmatrix}c_n^\dagger c_n.
\end{aligned}
\end{equation}
$\mathcal{H}_{xzy}(\mathbf{B})$ contributes a robust, field-insensitive background Chern vector $\mathbf{C}_\text{QAH}$, while the metallic component $\mathcal{H}_\text{metal}(\mathbf{B})$ provides the necessary degrees of freedom for Landau quantization under an applied magnetic field $\mathbf{B}$.
Here, the metallic bands are described by
\[
\begin{aligned}
\mathcal{H}_{\text{metal}}(\mathbf{B})={}&\sum_n\varepsilon_0c_n^\dagger c_n
+t_x e^{i\phi y}c_{n+\delta_x}^\dagger c_n\\
&+t_yc_{n+\delta_y}^\dagger c_n
+t_zc_{n+\delta_z}^\dagger c_n\\
&+\text{h.c.},
\end{aligned}
\]

and $ \mathcal{H}_{xzy}(\mathbf{B}) $ is identical to Eq.~(\ref{Eq1}), with the inclusion of an applied magnetic field $ \mathbf{B}$.
The magnetic field \( \mathbf{B} = (0, 0, B) \) is incorporated by introducing an extra flux \( \phi = a^2B/\phi_0 \) through the unit lattice area \( \delta_x \times \delta_y = a^2 \) \cite{LB1}.

As schematically illustrated in Fig. \ref{f4}(A)--(C), the coexistence of these bands enables the synthesis of a composite Chern vector $\mathbf{C} = (0, m, n)$. While the open Fermi surfaces of the 3D QAH surface states [Fig. \ref{f4}(B)] are immune to Landau quantization \cite{3DQHE1, 3DQHE2}, the closed Fermi surface of the metallic component facilitates the 3D quantum Hall effect (QHE) under a field $\mathbf{B} \parallel z$ \cite{3DQHCDW}. This allows the system to reach target states with different Chern vector magnitudes.

Numerical spectra and transport calculations in Fig. \ref{f4}(D)--(G) confirm that the Hall conductances $G_{xy}$ and $G_{yx}$ remain quantized according to the predicted formulas, even in the presence of inter-band coupling $t_d$. For instance, the system hosts stable $\mathbf{C} = (0, 1, 1)$ and $(0, 1, 2)$ phases at distinct energy windows, demonstrating the existence of the collective transport.

From a material perspective, the requirement for multiple coexisting orbitals is intrinsically satisfied in many candidates. Based on topological material databases, compounds such as RbSb, CsSb, RbAs~\cite{Candidate1}, and BiInN/Al$_2$O$_3$~\cite{Candidate2} are identified as promising candidates, as they naturally exhibit metallic bands nested within a nontrivial QAH gap. Through strategic heterostructure engineering or stacking, the proposed model offers a highly feasible route for experimental observation of controllable 3D Chern vectors.

\section{Discussion and conclusion}

Our work identifies a distinct quantized transport phenomenon that originates from the inherent chiral network manifested by the Chern vector $\mathbf{C} = (0, m, n)$. This network-based topology dictates unique collective properties and global connectivity.
Specifically, we demonstrate that the sum of transmission coefficients along both the \( x \)- and \( y \)-directions, rather than the individual ones, is protected by the Chern vector, preserving the quantized value \( (mL_y + nL_z) \). This summation relation manifests as observable quantized Hall conductances \( \pm(mL_y + nL_z) \).
A hallmark of these responses is their distinctive double Chern number components dependence, which serves as a unique experimental fingerprint. To facilitate experimental validation, we propose a class of candidate condensed-matter systems.

Beyond condensed-matter materials, the chiral surface states associated with $\mathbf{C} = (0, m, n)$ have been observed in photonic and acoustic crystals \cite{Chernvector, Chernvector2,Chernvector4}. Although directly measuring Hall conductances in these platforms remains challenging, the characteristic chiral network scattering [Figs.~\ref{f2}(D) and \ref{f2}(E)] can be validated by tracking wave-packet propagation. These distinct trajectories underpin the microscopic origin of the collective Hall conductances in Eqs.~(\ref{Hallyx}) and (\ref{Hallxy}).

In conclusion, our study advances the understanding of quantized transport in 3D topological systems. Notably, exploring the transport properties of Chern vectors with three non-zero components, $\mathbf{C} \!=\! (\mathcal{C}_{yz}, \mathcal{C}_{xz}, \mathcal{C}_{xy})$, represents an exciting avenue for future investigation, potentially unveiling even richer physics.

\begin{acknowledgments}
We appreciate the valuable discussions with Hong Yao, Li-Yuan Zhang, Ming Lu, Ming Gong, Qiang Wei and Rui-Chun Xiao.

This work was supported by the National Basic Research Program of China (Grants No. 2024YFA1409003 and No. 2022YFA1403700), National Natural Science Foundation of China (Grants No. 12350401 and No. 12204044).

\end{acknowledgments}

\section*{Data availability}
All essential data are available in the paper. Additional data are given in the supplementary file.

\section*{Supplementary data}
Further supporting data, including the codes for numerical calculations can be accessed via the GitHub website (\url{https://github.com/jianghuaphy-arch/chernvector}).

\appendix*
\section{Lattice model and Chern-vector characterization}

Here we provide the explicit lattice Hamiltonians and the calculation
of the corresponding Chern vector $
\mathbf{C} = (\mathcal{C}_{yz}, \mathcal{C}_{xz}, \mathcal{C}_{xy})
$  for the model introduced in
Eq.~(\ref{Eq1}).

The first sector, $\mathcal{H}_{xyz}$, is written in real space as
\begin{equation}
\begin{aligned}
\mathcal{H}_{xyz}={}&\sum_nT_0c_n^\dagger c_n\\
&+\sum_n\bigl(T_xc_{n+\delta_x}^\dagger c_n
+T_yc_{n+\delta_y}^\dagger c_n\\
&\qquad\quad+T_zc_{n+\delta_z}^\dagger c_n\bigr)+\mathrm{h.c.},
\end{aligned}
\end{equation}
where
\[
T_0= t\tau_z,
\qquad
T_x=\frac{M}{2}\tau_z-\frac{iA}{2}\tau_x,
\]
\[
T_y=\frac{M}{2}\tau_z-\frac{iA}{2}\tau_y,
\qquad
T_z=t_1\tau_z.
\]
Here, $\tau_{x,y,z}$ are Pauli matrices acting in the orbital space,
and $\delta_\alpha$ denotes the nearest-neighbor displacement along
the $\alpha$ direction.

Fourier transforming the real-space Hamiltonian gives
\begin{equation}
\begin{aligned}
\mathcal{H}_{xyz}(\mathbf{k})={}&t\tau_z+M\tau_z\cos k_x\\
&+A\tau_x\sin k_x+M\tau_z\cos k_y\\
&+A\tau_y\sin k_y+2t_1\tau_z\cos k_z.
\end{aligned}
\end{equation}

The second sector, $\mathcal{H}_{xzy}$, is obtained from
$\mathcal{H}_{xyz}$ by interchanging the roles of the $y$ and $z$
directions. Its momentum-space Hamiltonian is
\begin{equation}
\begin{aligned}
\mathcal{H}_{xzy}(\mathbf{k})={}&t\tau_z+M\tau_z\cos k_x\\
&+A\tau_x\sin k_x+2t_2\tau_z\cos k_y\\
&+M\tau_z\cos k_z+A\tau_y\sin k_z.
\end{aligned}
\end{equation}

The two sectors are coupled through
$T_d=t_d(\mathbf{I}_{2\times2}+\tau_x)$,
as introduced in the main text. For the numerical calculations, we take
$A=-t$, $M=2t$, $t_1=-0.3t$, and $t_2=-0.1t$,
with $t=1$ taken as the energy unit.  For these parameters, the bulk gap remains open for $t_d<0.5t$; the
evolution of the bulk spectrum with $t_d$ is shown in SM Note~1.

To characterize the topology, we calculate the Chern number
$\mathcal{C}_{\alpha\beta}(k_\gamma)$ of each two-dimensional momentum
slice at fixed $k_\gamma$ using the TKNN formula \cite{QH3}:
\begin{equation}
\begin{aligned}
\mathcal{C}_{\alpha\beta}(k_\gamma)
={}&-\frac{1}{\pi}\int\!\!\int_{\mathrm{BZ}}\sum_{E_c,E_v}\\
&\times\frac{\mathrm{Im}\!\left[\begin{aligned}
&\langle u_c|\partial_{k_\alpha}\mathcal{H}(\mathbf{k})|u_v\rangle\\[-2pt]
&\quad\times\langle u_v|\partial_{k_\beta}\mathcal{H}(\mathbf{k})|u_c\rangle
\end{aligned}\right]}{(E_c-E_v)^2}\\
&\times\,dk_\alpha\,dk_\beta,
\end{aligned}
\end{equation}
where $\alpha,\beta,\gamma\in\{x,y,z\}$, and $|u_c\rangle$
($|u_v\rangle$) denotes an eigenstate in the conduction (valence)
band with energy $E_c>0$ ($E_v<0$).

The system remains gapped throughout the
Brillouin zone, and the corresponding slice Chern numbers are
independent of the transverse momentum for all
$k_{x,y,z}\in[-\pi,\pi]$.  The resulting Chern vector is
\[
\mathbf{C}
=
(\mathcal{C}_{yz},\mathcal{C}_{xz},\mathcal{C}_{xy})
=
(0,1,1).
\]
This Chern vector remains unchanged for finite inter-sector coupling
$t_d$ as long as the bulk gap remains open. At $t_d\simeq0.5t$, the
bulk gap closes, signaling a topological phase transition
[see SM Note~1].

\bibliography{references}

\end{document}